\documentclass[aps,prd,showpacs,eqsecnum,twocolumn,superscriptaddress]{revtex4}

\usepackage{latexsym} \usepackage{amssymb} \usepackage{amsfonts}
\usepackage{amsmath} \usepackage{bm} \usepackage[dvips]{graphicx}
\usepackage{color}
\usepackage{subfigure} \usepackage{times} \usepackage{units}
\usepackage{hyperref}

\usepackage[utf8x]{inputenc} \usepackage{amssymb,amsmath}
\usepackage{graphicx} \usepackage[squaren]{SIunits} 

\begin{document}
\title{Total-variation-based methods for gravitational wave denoising}
\author{Alejandro \surname{Torres}}\affiliation{Departamento de
  Astronom\'{\i}a y Astrof\'{\i}sica, Universitat de Val\`encia,
  Dr. Moliner 50, 46100, Burjassot (Val\`encia), Spain} 

\author{Antonio \surname{Marquina}}\affiliation{Departamento de
  Matem\'atica Aplicada, Universitat de Val\`encia,
  Dr. Moliner 50, 46100, Burjassot (Val\`encia), Spain}  
  
\author{Jos\'e A. \surname{Font}}\affiliation{Departamento de
  Astronom\'{\i}a y Astrof\'{\i}sica, Universitat de Val\`encia,
  Dr. Moliner 50, 46100, Burjassot (Val\`encia), Spain}
  \affiliation{Observatori Astron\`omic, Universitat de Val\`encia, C/ Catedr\'atico 
  Jos\'e Beltr\'an 2, 46980, Paterna (Val\`encia), Spain}
  
\author{Jos\'e M. \surname{Ib\'a\~nez}}\affiliation{Departamento de
  Astronom\'{\i}a y Astrof\'{\i}sica, Universitat de Val\`encia,
  Dr. Moliner 50, 46100, Burjassot (Val\`encia), Spain}   
    \affiliation{Observatori Astron\`omic, Universitat de Val\`encia, C/ Catedr\'atico 
  Jos\'e Beltr\'an 2, 46980, Paterna (Val\`encia), Spain}


\begin{abstract}
We describe new methods for denoising and detection of gravitational waves embedded in additive Gaussian noise. 
The methods are based on Total Variation denoising algorithms. These algorithms, which do not need any a priori 
information about the signals, have been originally developed and fully tested in the context of image processing. 
To illustrate the capabilities of our methods we apply them to two different types of numerically-simulated gravitational 
wave signals, namely {\it bursts} produced from the core collapse of rotating stars and waveforms from binary black hole mergers. We explore the parameter space of the methods to find the set of values best suited for denoising gravitational 
wave signals under different conditions such as waveform type and signal-to-noise ratio. Our results show that noise 
from gravitational wave signals can be successfully removed with our techniques, irrespective of the signal morphology 
or astrophysical origin. We also combine our methods with spectrograms and show how those can be used simultaneously 
with other common techniques in gravitational wave data analysis to improve the chances of detection.
\end{abstract}

\pacs{
04.30.Tv,	
04.80.Nn,	
05.45.Tp,	
07.05.Kf,	
02.30.Xx.	
}
\maketitle

\section{Introduction}
\label{section:intro}

After a sustained effort spanning several decades 
gravitational wave (GW) astronomy is expected to 
become a reality in the next few years. 
The next generation of ground-based, 
laser interferometer gravitational wave detectors, 
Advanced LIGO~\cite{LIGO},  Advanced Virgo~\cite{Virgo} 
and KAGRA~\cite{KAGRA}, are presently being upgraded and 
will begin operating before the end of this decade. 
The significant improvement in sensitivity with respect to 
previous detectors will make possible the direct detection 
of GWs in a frequency band ranging from about 
10 Hz to a few kHz from a large class of sources including, among others, 
isolated spinning neutron stars, 
coalescing binary neutron stars and/or (stellar mass) black holes,  
core collapse supernovae and gamma-ray bursts, 
each one of them with a characteristic signal waveform.

One of the most challenging problems in signal data analysis 
(and GW data analysis in particular) is noise removal. 
Almost every process of transmission, detection, 
amplification or processing, add random distortions to the original signal.
Noise in GW interferometers is particularly problematic 
because the signals from most sources are in the limit of detectability. 
Additionally, real signals will be disturbed by noise transients and 
noise frequency lines will appear often in the detector data.  
These transients are often indistinguishable from the true signals, 
hence algorithms to identify noise artifacts and to ``veto''  
them are required \cite{Parameswaran:2014, credico:2005}. 

GW data analysis algorithms have had a 
great development over the past decade 
(see~\cite{Jaranowski2012} and references therein). 
Specific analysis techniques have been developed 
according to the specific type of source~\cite{Abbott:2009} 
and deterministic signals exist for rotating neutron stars 
(continuous signals), coalescing binaries (transient, modelled signals), 
and supernova explosions (transient, unmodelled signals). 
For coalescing compact binaries the inspiral part of the (chirp) signal 
can be detected by correlating the data with 
analytic waveform templates and maximizing such correlation 
with respect to the waveform parameters. 
When the signal-to-noise ratio (SNR) of the filter output 
over a wide bandwith of the detector exceeds an 
optimal threshold,  the matched-filtering technique 
generates a trigger associated with a specific template. 

On the other hand, the detection of the continuous 
gravitational wave emission from radio pulsars and 
spinning neutron stars is unpractical for 
matched-filtering techniques due to the exceedingly 
large computational resources it would require. 
Such continuous GW signal is weak and only galactic 
sources (within a few hundred pc) are expected to be detected. 
 The GW emission from millisecond pulsars is stable and allows 
for coherent signal integration for long time intervals. 
Data analysis methods designed for continuos GWs 
differ by duration and sky coverage (all-sky or targeted) \cite{abadie:2012,prix:2009}. 
Computational requirements from long time all-sky search 
have led to the Einstein@Home initiative~\cite{einstein-home}.
 
The aspherical gravitational collapse of the core of massive stars 
produces a short duration signal with a significant power 
in the kHz frequency band. Unlike coalescing compact binaries, 
such supernova ``burst'' GW signals are unmodelled 
due to uncertainties with the large number of parameters involved. 
Numerical simulations are computationally expensive which renders 
impractical to produce a comprehensive enough template bank 
against which employ matched-filtering techniques for detection. 
Time-frequency analysis is used instead \cite{anderson:1999} in which the signal 
is decomposed in its frequency components via Fourier transform 
or using wavelets, and a trigger is generated if some component 
is above the detector baseline noise. 
This approach involves the analysis of the coincidences 
among the various detectors, studying the triggers either directly, 
using cross-correlation methods or, in a more general way, 
using coherent methods \cite{bose:2000,klimenko:2008}.

For GW burst signals in particular, Bayesian inference methods 
have been proposed to extract physical parameters and reconstruct 
signal waveforms from noisy data~\cite{Rover:2009,Engels:2014}. 
In practice a direct solution of the normal equations derived from the
associated least squares problems can lead to numerical difficulties
when the matrix of the system is large, dense and close to singular.
These difficulties can be addressed using the technique of 
{\it singular value decomposition} (SVD) that reduces 
the complexity and singularity of the problem by means of 
computing a small 
number of basis vectors, i.e.~an orthogonal set of eigenvectors 
derived from numerical relativity simulated waveforms. 
Markov Chain Monte Carlo (MCMC) algorithms fit the basis vectors 
to the data, providing waveform reconstruction with confidence 
intervals and distributions of physical parameters, assuming that 
the principal components are related to astrophysical properties 
of the source. The use of a good Bayesian prior often makes 
the problem computationally complex and ill-conditioned resulting in
an over-fitted signal recovery.  

In this paper we deviate from the standard approaches used 
in the GW data analysis community by assessing a method based on 
Total Variation (TV) norm regularized algorithms for 
denoising and detection of GWs 
embedded in additive Gaussian noise. 
Our motivation comes from the idea of adding a regularization term
to the error function (fidelity), weighted by a positive Lagrange multiplier, 
in order to control the over-fitting, where the Lagrange multiplier 
measures the relative importance of the data-dependent fidelity term. 
TV-norm regularization was introduced in 1992 in ~\cite{Rudin:1992} for
denoising problems. This regularization becomes successful because it
uses the regularization strategy based on a $\text{L}_1$-norm.
TV-denoising is based on such regularization and the variational
model is called ROF model, after Rudin, Osher and Fatemi. It is one
of the most common methods employed for image denoising since it avoids 
spurious oscillations and Gibbs phenomenon near edges.
Since the publication of compressed sensing (CS) 
methods~\cite{Donoho:2006}, based on the $\text{L}_1$-norm 
regularization which allows to accurately reconstruct a 
signal or image from a small part of the data in a very efficient way, 
this technique has been employed with very successful 
results in medical imaging, radar imaging, and magnetic 
resonance imaging (see ~\cite{Goldstein:2009} and references therein).  
The main advantage of the $\text{L}_1$-norm regularization is that
it favors {\it sparse} solutions, i.e.~very few nonzero components of the 
solution or its gradient. In addition the algorithm to find $\text{L}_1$-norm
minimizers is extremely efficient despite this norm is not differentiable.
Split Bregman method was developed by Goldstein and Osher~\cite{Goldstein:2009} 
to efficiently solve most common $\text{L}_1$-regularized problems 
and it is particularly effective to solve denoising problems based
on the ROF model.  

More precisely, in this work we apply two different TV-denoising techniques
in the context of GW denoising. Arguably, the main advantage of these techniques 
is that no a priori information about the astrophysical source 
or the signal morphology is required to perform the denoising. 
As we illustrate below, this main feature allows us to obtain 
satisfactory results for two different catalogs of 
gravitational waveforms comprising signals with very different structure. 
Our aim is to find optimal values of the parameters of the 
ROF model that can assure a proper noise removal. 
In order to do so  we modify the ROF problem to take into 
consideration the sensitivity curve from the Advanced LIGO detector. 
We restrict the formulation of the problem to 1D, 
since the available gravitational wave catalogs 
used are one-dimensional, even though the algorithm 
can be easily extended to higher dimensions. 
We emphasize that there are no restrictions about the data, 
and in this way the denoising can be performed in both the time 
or the frequency domain. 

This paper is organized as follows:
Section \ref{section:overview} explains total variation methods and 
algorithms. Section \ref{section:gw} describes 
the gravitational waveform catalogs employed to assess our methods. 
In Section \ref{section:parameter_estimation} 
we adapt the general problem to the specific case of 
GW signals and detector noise and we obtain a satisfactory
value of the regularization parameter for both algorithms. 
In Section \ref{section:application} we discuss the results 
of applying the methods to signals from the two GW 
catalogs either when they are applied in a standalone fashion or 
in combination with other common tools in data analysis such as
spectrograms. Finally, the conclusions of our work are presented 
in Section \ref{section:summary}.  Technical details regarding noise
generation are briefly considered in the appendix.

\section{Total variation based denoising methods}
\label{section:overview}
\subsection{Variational models for denoising: 
Total variation based regularization}

We shall assume the general linear degradation model 
\begin{equation}
\label{eq:denoising_problem}
f=u+n~,
\end{equation}
where $f$ is the observed signal, $n$ is the noise and $u$ is 
the signal to be recovered. For the sake of understanding of
the adopted mathematical models we assume that $n$ is
Gaussian white noise, (i.e.~$n$ is a square integrable
function with zero mean). Moreover, throughout this section we 
will assume that the signals belong to a $k$-dimensional 
Euclidean space provided with discrete $\text{L}_1$ and $\text{L}_2$  
norms.

The problem of signal denoising
consists of estimating a (clean) signal $u$
whose square of the $L^2$-distance to 
the observed noisy signal $f$ 
is the variance of the noise, i.e. 
\begin{equation}\label{eq:fidelity_term}
||u-f||_{\text{L}_2}^2=\sigma^2~.
\end{equation}
where $\sigma$ denotes the standard deviation of the noise.

Classical models and algorithms for solving the 
denoising problem are based on least squares, 
Fourier series and other $\text{L}_2$ norm approximations. 
A least squares problem can be solved by computing the solution 
of the associated normal equations, 
which is a linear system of equations 
where the unknowns are the coefficients of a 
linear combination of polynomials or a 
wavelet basis~\cite{Irani:1993}. 
The main drawback of this technique is that the results 
are contaminated by Gibbs' phenomena (ringing) 
and/or smearing near the edges (see \cite{Marquina:2009} 
and references therein). Moreover, the linear system to be solved
is large, (related to the size of the sample of the observed signal $f$)
and ill-conditionned, i.e.~close to singular.

The usual approach to overcome 
these problems is to regularize
the least squares problem using 
an auxiliary energy (`prior') 
$R(u)$, and solve the following 
constrained variational problem
\begin{eqnarray}
\underset{u} {\min} \quad R(u)&&\\\nonumber
 \mbox{subject to}&& ||f-u||^2_{L^2}=\sigma^2
\end{eqnarray}
where the functional $R(u)$
measures the quality of the signal $u$ in the sense that smaller 
values of $R(u)$ correspond to better signals. 
This general model can be applied to 1D signals,
2D images or multidimensional volume data.

The above variational problem has a 
unique solution when the energy $R(u)$ is convex. 
We will assume from now on that $R(u)$ is convex. 
The constrained variational problem can be formulated 
 as an unconstrained variational problem using  the Tikhonov regularization.
The regularization consists of  adding the constraint ("fidelity term") 
weighted by a positive Lagrange multiplier $\mu>0$ 
(also unknown) to the energy $R(u)$
\begin{equation}
 \label{eq:unconstrainL2}
 u=\underset{u} {\text{argmin}}\left\{R(u)+\frac{\mu}{2} \, ||f-u||_{\text{L}_2}^2\right\}~.
 \end{equation}
There exists a unique value of $\mu>0$ such that the unique solution $u$ 
matches the constraint. The Lagrange multiplier $\mu > 0$ becomes a 
scale parameter in the sense that larger values of $\mu$ allow to recover
finer scales in a scale space determined by the regularizer functional $R(u)$. 
It can be understood as that $\mu > 0$ controls the relative
importance of the fidelity term.

If $R(u):=\int{|\nabla u|^2}$ where the integral is extended 
to the domain of the signal either discrete or continuous,
then the model (\ref{eq:unconstrainL2}) 
becomes the so-called Wiener filter.
In order to compute the solution in this case we solve
the associated Euler-Lagrange  equation 
\begin{equation}
\label{eq:EL_Wiener}
\Delta u + \mu (f-u) = 0~,
\end{equation}
under homogeneous Neumann boundary conditions.
In the previous equation and in the definition of the energy $R(u)$, 
$\Delta$ and $\nabla$ stand, respectively, for the (discrete) Laplacian 
and gradient operators. This equation corresponds to a nondegenerate second 
order linear elliptic differential equation,  
which is easy to solve due to differentiability and strict convexity of the energy term. 
We note that this equation satisfies the conditions that guarantee uniqueness of the
solution (see~\cite{isa:2009} and references therein).
Indeed, the equation can be efficiently solved 
by means of the Fast Fourier Transform (FFT). 
The choice of a quadratic energy for the regularizer makes the variational
problem more tractable. It encourages Fourier coefficients of the
solution to decay towards zero, surviving the ones representing 
the processed signal $u$.
However this good behavior is no longer valid when noise
is present in the signal. Noise amplifies high frequencies
and the recovered smooth solution $u$ prescribed by the model
contains spurious oscillations near steep gradients or edges
(see \cite{Marquina:2000, Voguel:1998,Osher:2005}).   

The Wiener filter procedure reduces noise 
by shrinking Fourier coefficients of the signal towards zero but adds 
spurious oscillations due to the Gibbs' phenomena.

In order to avoid the  aforementioned problems arising by using
quadratic variational models, Rudin, Osher and Fatemi proposed in \cite{Rudin:1992} 
the TV norm as regularizing functional for the variational 
model for denoising (\ref{eq:unconstrainL2})
\begin{equation}\label{tv}
TV(u)=\int |\nabla u|
\end{equation}
where the integral is defined on the domain of the signal.
The ROF model consists of solving the variational problem for denoising:
\begin{equation}
\label{eq:rof}
u=\underset{u} {\text{argmin}}\left\{TV(u)+\frac{\mu}{2}||u-f||_{\text{L}_2}^2\right\}~.
\end{equation}
The TV norm energy is essentially the $\text{L}_1$-norm 
of the gradient of the signal. 
Although  many $\text{L}_1$ based norms have been usually 
avoided because of its lack of differentiability,  
the way the $\text{L}_1$-norm is used in the ROF model 
has provided a great success in denoising problems. The
ROF model allows to recover edges of the original signal
removing noise and avoiding ringing.
The parameter $\mu>0$ runs a different scale space as in the Wiener model.
Since the energy is convex there is a unique optimal value of the 
Lagrange multiplier $\mu>0$ (scale) for which equation 
(\ref {eq:fidelity_term}) is satisfied.
When the standard deviation of the noise is unknown a heuristic 
estimation of $\mu$ is needed to find the optimal value.
Indeed, if we choose a large value of $\mu$ the ROF model
will remove very little noise, while finer scales will be destroyed
if small values of $\mu$ are chosen instead.

The use of $\text{L}_1$-norm related energies for least squares
regularization has been popularized following
the pioneering contribution 
of Rudin, Osher and Fatemi. 
For example, soft thresholding is a denoising algorithm
related to $\text{L}_1$-norm minimization introduced by Donoho
in \cite{donoho:1994,donoho:1995}. 
The $\text{L}_1$-norm regularization selects a unique
{\it sparse} solution, i.e., solutions with few nonzero elements. This 
property is essential for {\it compressive sensing} problems, (see 
\cite{Donoho:2006, Candes_Romberg_Tao:2006}). 
An earlier application of penalizing with the $\text{L}_1$-norm
is the LASSO regression proposed in
the seminal paper by R. Tibshirani in \cite{lasso}.

Since the ROF model uses the TV-norm 
the solution is the only one with the {\it sparsest} gradient. Thus, the
ROF model reduces noise by {\it sparsifying} the gradient of the signal and
avoiding spurious oscillations (ringing).

\subsection{Algorithms for TV based denoising: Split Bregman Method}

We shall present the algorithms we will use for the computation
of the solution of the ROF model.

The standard method to solve nonlinear smooth optimization problems
is to compute the solution of the associated Euler-Lagrange equation.
The ROF model consists of a nonsmooth optimization problem and the
associated Euler-Lagrange equation can be expressed as
\begin{equation}
\label{eq:EL_TV}
\nabla \cdot \frac{\nabla u }{|\nabla u|}+ \mu (f-u) = 0~,
\end{equation}
where the differential operator becomes singular and
 has to be defined properly when $|\nabla u| = 0$ (see \cite{Meyer: 2001}). 

The first algorithm we will use  is the regularized ROF algorithm (rROF hereafter).
This algorithm  computes an approximate 
solution of the ROF model by smoothing the total variation energy.
Since the Euler-Lagrange derivative of the TV-norm is not well defined
at points where $\nabla u=0$, the TV functional is slightly perturbed as
\begin{equation}
TV_{\beta}(u):= \int \sqrt{|\nabla u|^2+\beta}~,
\end{equation}
where $\beta$ is a small positive parameter.
We will use the expression
\begin{equation}
\int |\nabla u|_\beta
\end{equation}
with the notation
\begin{equation}
|v|_\beta=\sqrt{|v|^2+\beta}
\end{equation}
for $v\in \Re^p$ where $p$ is dimension of the signal.

Then the rROF model in terms of the small positive 
parameter $\beta>0$  reads as
\begin{equation}
\label{eq:rof_beta}
u=\underset{u} {\text{argmin}}\left\{TV_{\beta}(u)+\frac{\mu}{2}||u-f||_{\text{L}_2}^2\right\}~,
\end{equation}
and the associated Euler-Lagrange equation will be
\begin{equation}
\label{eq:EL_TV_beta}
\nabla \cdot \frac{\nabla u }{|\nabla u|_{\beta}}+ \mu (f-u) = 0~.
\end{equation}

Assuming homogeneous Neumann boundary conditions 
Eq.~(\ref{eq:EL_TV_beta}) 
becomes a nondegenerate second order nonlinear elliptic 
differential equation whose solution is smooth.
In order to solve the above equation we use 
conservative second order central differences for the
differential operator and point values for the source term.
The approximate  solution will be obtained by means
of a nonlinear Gauss-Seidel iterative procedure that uses as
initial guess the observed signal $f$. 
We apply homogeneous Neumann boundary values to ensure
conservation of the mean value of the signal.

The second algorithm we shall use is the so-called ``Split Bregman Method"
(SB hereafter) proposed in~\cite{Goldstein:2009}. 
The method consists of an iterative alternating procedure that splits
the approximation of the minimizer into two
steps: first, solving the least squares minimization and second, 
performing 
direct minimization of the TV energy 
using the ``shrinkage function" and freezing the fidelity term
computed at the approximation obtained in the first step.

The splitting process  is  combined with the Bregman iterative
refinement \cite{Bregman:1967}.
The Bregman iterative procedure can be applied to a general
fidelity term. Let us assume that $E(u)$ is a nonnegative convex energy and
we wish to solve the following constrained variational problem:
\begin{eqnarray}
\underset{u} {\min} \quad E(u)&&\\\nonumber
 \mbox{subject to}&& Au=b
\end{eqnarray}
where $A$ is some linear operator and $b$ is a vector.
We rewrite the variational problem as an unconstrained optimization by
introducing a Lagrange multiplier $\lambda>0$ that weights the influence of the
fidelity term as
\begin{equation}
\label{eq:unconstrained3}
u=\underset{u} {\text{argmin}}\left\{E(u)+\frac{\lambda}{2}||b-Au||_{\text{L}_2}^2\right\}~.
\end{equation}
If we choose small $\lambda>0$ then the solution 
of the variational problem does not
accurately enforce the constraint. 
The solution we need is to let $\lambda$ large.
Alternatively, we shall use the following Bregman iterative 
refinement to enforce 
the constraint $Au=b$ accurately 
using a fixed small value for $\lambda$:
\begin{eqnarray}
u^{k+1}&=&\underset {u} {\text{argmin}}~\left\{E(u)+\frac{\lambda}{2}||b^k-Au||^2_{\text{L}_2}\right\}\,,\label{eq:bregmansimply1} \\
b^{k+1}&=&b+b^k-Au^{k+1}\,.\label{eq:bregmansimply2}
\end{eqnarray}
Roughly speaking we add the residual error (of the fidelity term) back to the constraint
to solve a new variational problem in each iteration. 

Next we shall sketch the SB method
for the particular case of the ROF model
applied to the one dimensional signals 
that we will use in our numerical experiments.
The SB method to solve the ROF model combines the 
Bregman iterative procedure  
described by (\ref{eq:bregmansimply1}) and  
(\ref{eq:bregmansimply2}) with the decoupling of the 
TV variational problem 
into $\text{L}_1$ and $\text{L}_2$ 
portions of the energy to be minimized.
Each ROF problem appearing in every Bregman iteration is solved by splitting
the $L^1$ term and $L^2$ terms and minimizing them separately.

The procedure reads as follows. 
We solve each ROF problem in the iterative procedure
(\ref{eq:bregmansimply1}) by
introducing a new variable $d$, and we replace
$\nabla_x u$ by $d$ and we add the constraint
$\nabla_x u = d$, where $\nabla_x u$
represents the one-dimensional gradient.
We set the notation
$$s(b,u,d):=||b+(\nabla_x u) - d||_{\text{L}_2}^2\,.$$
Then, we formulate the following Bregman iterative procedure applied on
the new constraint (see \cite{Goldstein:2009}):
\begin{equation}
 (u^{k+1},d^{k+1})=\underset{u,d} {\text{argmin}} \left\{|d|+ \frac{\mu}{2}||f -  u||_{\text{L}_2}^2+\frac{\lambda}{2} s(b^k,u,d)\right\}\label{eq:bregmansimply3}
\end{equation}
\begin{equation}
\qquad b^{k+1}=b^k+(\nabla_x u^{k+1}) - d^{k+1}\,,\,\,\,\qquad\qquad\label{eq:bregmansimply4}
\end{equation}
where we set $s(b^k,u,d):=b^k+(\nabla_x u) - d$.

Thus, we can iteratively minimize with respect to $u$ and $d$ separately
and the split Bregman iterative procedure will read as follows
\begin{eqnarray}
\label{split_bregman_iteration1}
 u^{k+1} &=&\underset {u} {\text{argmin}} \left\{\frac{\mu}{2}||f-u||_{\text{L}_2}^2
  + \frac{\lambda}{2} s(b^k,u,d^k)\right\} \\
\label{split_bregman_iteration2} 
 d^{k+1} &=&\underset {d} {\text{argmin}}~|d|+\frac{\lambda}{2}||b^k + (\nabla_x u^{k+1}) - d||_{\text{L}_2}^2\\ 
\label{split_bregman_iteration3}
 b^{k+1}&=&b^k+(\nabla_x u^{k+1}) - d^{k+1}\,.
\end{eqnarray}

Since the two parts are decoupled, they can be solved independently. 
The energy of the first step is smooth (differentiable) and it 
can be solved using common techniques such as the Gauss-Seidel method. 
On the other hand, $d$ can be computed to optimal values that solve 
the problem using shrinkage operators,
\begin{equation}
\label{eq:shrink}
d^{k+1}={\text{shrink}}(b^k+ (\nabla_x u^{k+1}),1/\lambda)~,
\end{equation}
\begin{equation}
\label{eq:shrink2}
{\text{shrink}}(x,\gamma)=\frac{x}{|x|}\ast \text{max}(|x|-\gamma,0)~.
\end{equation}

We only use one iteration for the splitting steps and the final algorithm only consists
of just one loop (see \cite{Goldstein:2009} for a detailed discussion). 

The SB algorithm we will use is written as follows:
\begin{itemize}
\item Initial guess: $u^0=f$, $d^0=0$ and $b^0=0$
\item while $||u^k-u^{k-1}||_{\text{L}_2}>\text{tol}$
\begin{itemize}
\item $u^{k+1}=G^k$ 
\item $d^{k+1}={\text{shrink}}(b^k+ (\nabla_x u^{k+1}),1/\lambda)$
\item $b^{k+1}=b^k+(\nabla_x u^{k+1}) - d^{k+1}$
\end{itemize}
\item end
\end{itemize}
where the Gauss-Seidel step can be expressed as the loop
\begin{itemize}
\item for $j$
\begin{itemize}
\item $G^k_j=\displaystyle\frac{\lambda}{\mu+2 \lambda} (u^k_{j+1}+u^k_{j-1}-(d^k_{j}-d^k_{j-1})+(b^k_{j}-b^k_{j-1}))+\frac{\mu}{\mu+2 \lambda} f_j$
\end{itemize}
\item end $j$
\end{itemize}
and $(\nabla_x u^{k+1})_j=u^{k+1}_{j+1}-u^{k+1}_{j}$
where $j$ runs the component positions of the discretization.

\section{Gravitational waves catalogs}
\label{section:gw}

We turn next to describe the main features of the signals 
included in the two GW waveform catalogs
we use to assess the two methods we have just presented.

\subsection{Rotating core collapse catalog}

At the final phase of their evolution, massive stars in the
9-40 M$_{\odot}$ range develop iron-group element
cores which are dynamically unstable against gravitational collapse. 
According to the standard model of type II/Ib/Ic supernovae, 
the collapse is initiated by electron captures and
photo-disintegration of atomic nuclei when the iron core
exceeds the effective Chandrasekhar mass. 
As the inner core reaches nuclear density and 
the equation of state (EoS) stiffens,
the collapse stops and is followed by the bounce of the inner core.
A strong shock wave appears in the boundary between the inner core 
and the supersonically infalling outer core. 
Numerical simulations try to elucidate if this shock wave is 
powerful enough to propagate from the outer core and
across the external layers of the star.
In the most accepted scenario, energy deposition by neutrinos,
convective motions, and instabilities in the standing shock wave, 
together with general-relativistic effects, are necessary elements 
leading to successful explosions. 

In the core collapse scenario, conservation of angular momentum
makes rotating cores with a period of one second to produce
millisecond period proto-neutron stars, with a rotational energy
of about $10^{52}$ erg.
The bulk of gravitational radiation is emitted during bounce,
when the quadrupole moment changes rapidly, which produces
a burst of gravitational waves with a duration of about 10 ms and
a maximum (dimensionless) amplitude of about $10^{-21}$ at
a distance of $10$ kpc.
Broadly speaking, GW signals from this mechanism exhibit
a distinctive morphology characterized by a steep rise
in amplitude to positive values before bounce followed
by a negative peak at bounce and a series of damped oscillations
associated with the vibrations of the newly formed proto-neutron star
around its equilibrium solution. 

\begin{figure}
        \centering
        {\includegraphics[width=7.0cm]{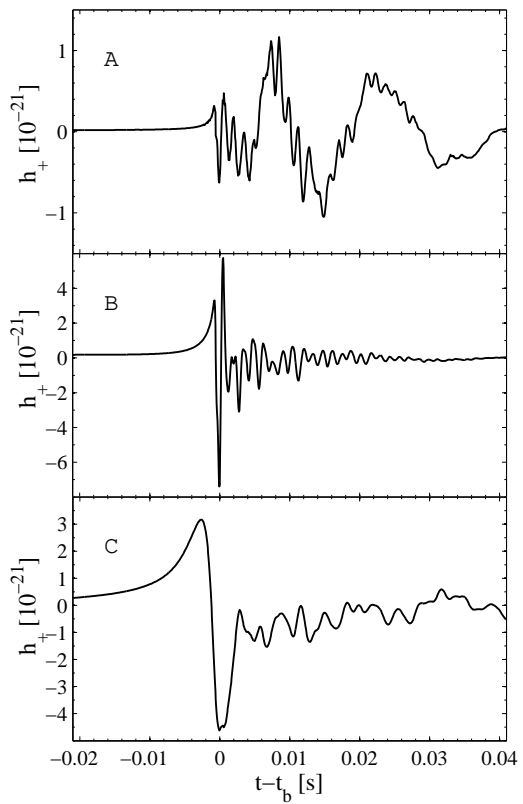}} 
        \caption{Gravitational waveforms of three representative signals from the core collapse catalog of~\cite{Dimmelmeier:2008} with different values of the degree of differential rotation. The equation of state and the progenitor mass are fixed for all represented signals. Signal ``s20a1o05\_shen'' (A) is show in the upper panel, signal ``s20a2o09\_shen'' (B) is shown in the middle panel, and signal ``s20a3o15\_shen'' (C) in the bottom panel. $t_{\rm b}$ indicates the time of bounce.}
         \label{fig:signals}
\end{figure}

Catalogs of gravitational waveforms from core collapse supernovae
have been obtained through numerical simulations 
with increasing realism in the input physics 
(see~\cite{fryer} and references therein). 
In our study, and for illustrative purposes, we employ only
the catalog developed by Dimmelmeier et al.~\cite{Dimmelmeier:2008},
who obtained 128 waveforms from general relativistic simulations of
rotating stellar core collapse to a neutron star.
The simulations were performed with the CoCoNuT code and
include a microphysical treatment of the nuclear EoS,
electron capture on heavy nuclei and free protons,
and an approximate deleptonization scheme
based on spherically symmetric calculations
with Boltzmann neutrino transport.
The simulations considered two tabulated EoS, those 
of~\cite{Shen:1998} and~\cite{Lattimer:1991} and
a wide variety of rotation rates and profiles and progenitor masses. 

The morphology and temporal evolution of the waveform signals
of the catalog of~\cite{Dimmelmeier:2008} are determined by
the various parameters of the simulations.
The initial models include solar metallicity,
non-rotating progenitors with masses at zero age main sequence of
$11.2~M_\odot$, $15.0~M_\odot$, $20.0~M_\odot$ and $40.0~M_\odot$.
Rotation has a strong effect on the resulting waveforms
and it is fixed by the precollapse central angular velocity
which is set to values from
$\Omega_{\text{c,i}}=0.45~\text{to}~13.31~\text{rad}~\text{s}^{-1}$.
The angular velocity of the models is given by
\begin{eqnarray}
\Omega=\Omega_{\text{c,i}}\frac{A^2}{A^2+r^2\sin^2\theta}\,,
\end{eqnarray}
$r\sin\theta$ being the distance to the rotation axis,
and $A$ being a length parameterizing the degree of differential rotation.
The specific values used in the simulations are
$A=50,000$ km (almost uniform rotation),
$A=1,000$ km, and $A=500$ km (strong differential rotation).
The GW amplitude maximum, $|h_{\rm max}|$, is proportional 
to the ratio of rotational energy to gravitational energy
at bounce, $\frac{T_b}{|W|_b}$.
The gravitational waveforms of the catalog are computed employing
the Newtonian quadrupole formula where the maximum dimensionless
gravitational wave strain is related to the wave amplitude $A_{20}^{\text{E2}}$ by
\begin{equation}
h=\frac{1}{8}\sqrt{\frac{15}{\pi}}\frac{A_{20}^{\text{E2}}}{D}=8.8524\times10^{-21}\frac{A^{\text{E2}}_{20}}{10^3~\text{cm}}\frac{10~\text{kpc}}{D},
\end{equation}
where $D$ is the distance to the source.

We focus on three representative signals from the catalog
to assess our algorithms.
These three waveforms, shown in Fig.~\ref{fig:signals},
cover the signal morphology of the catalog,
as explained in  \cite{logue:2012}.
These signals are labelled ``s20a1o05\_shen'',
``s20a2o09\_shen'', and  ``s20a3o15\_shen''
in the original catalog of~\cite{Dimmelmeier:2008}.
We rename them, respectively, as signals ``A", ``B", and ``C",
in the following, to simplify the notation.
The three values of the degree of differential rotation produce,
in particular, the most salient variations in the waveform morphology. 

\subsection {Binary black holes catalog}

%
 \begin{figure}
  \centering
    \includegraphics[width=7.0cm]{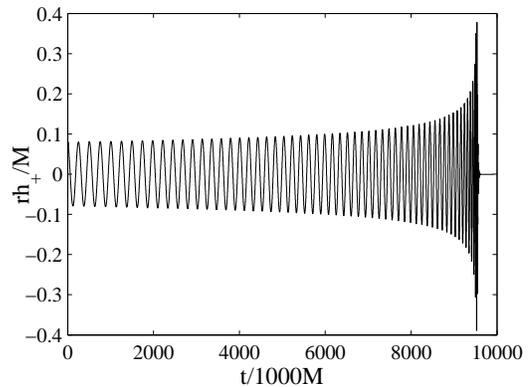}
  \caption{Gravitational waveform of a representative signal from the BBH catalog of~\cite{Mroue:2013}, signal ``0001", used for our test purposes.}
  \label{fig:bbh_signal}
\end{figure}

The second type of GW signals we use for our study
is that from the inspiral and merger of binary black holes.
Along with the merger of binary neutron stars,
such events are considered the most promising sources
for the first direct detection of gravitational waves.
This kind of systems evolve through three distinctive phases,
inspiral, merger, and ringdown, each one of them giving rise
to a different signal morphology.

During the inspiral phase, the orbital separation
between the two compact objects decays due to GW emission.
Analytic approximations of general relativity such as
post-Newtonian expansions or the Effective One Body method
describe to high accuracy the motion of the system and
the radiated GW content during such inspiral phase.
These were, until recently, the only approaches available
to model the GW signal before the merger of the compact binary.
After the merger, on the other hand, the resulting single 
BH asymptotically reaches a stationary state characterized
by the ringdown of its quasi-normal modes of oscillation,
whose waveform signal can be computed using standard techniques
from BH perturbation theory.
Either of these techniques, however,
cannot be used to model the signal during the merger phase
when the peak of the gravitational radiation is produced.
At this phase, the signal waveform has to be computed solving
the full Einstein equations with the techniques of numerical relativity,
a long-lasting challenging problem that was only recently
finally solved~\cite{pretorius:2005,campanelli:2006,baker:2006}.
Since those breakthroughs many numerical relativity simulations of
BBH mergers followed which led to the first BBH waveform catalogs
of the entire signal 
(late inspiral, merger, and quasi-normal mode ringdown)
based solely on fully numerical relativity approaches.

Nowadays, numerical relativity waveforms for BBH mergers
have become increasingly more accurate and 
span the entire seven-dimensional parameter space,
namely initial spin magnitudes, angles between the initial spin vectors
and the initial orbital angular momentum vector,
angles between the line segment connecting
the centers of the black holes and the initial spin vectors
projected onto the initial orbital plane,
mass ratio, number of orbits before merger (late inspiral), 
initial eccentricity, and final spin.
State-of-the-art waveforms are in particular reported 
by Mrou\'e et al \cite{Mroue:2013},
and these are the ones we use for our tests.
This catalog includes 174 numerical simulations
of which 167 cover more than 12 orbits
and 91 represent precessing binaries.
It also extends previous simulations to large mass ratios
(from 1 to 8) and includes simulations with the first systematic
sampling of eccentric BBH waveforms.
In addition, the catalog incorporates new simulations
with the highest BH spin studied to date ($0.98$).
As for the case of the core collapse burst catalog,
it is sufficient to focus on a representative signal
from the BBH catalog in order to illustrate
the performance of our denoising algorithms.
To such purpose we select signal labelled ``0001" 
from the Mrou\'e et al \cite{Mroue:2013} catalog, 
which is shown in Fig.~\ref{fig:bbh_signal}.

\section{Regularization parameter estimation}
\label{section:parameter_estimation}

Denoising results are strongly dependent on the value of the regularization parameter $\mu$. The optimal value of  $\mu$ 
that produces the best results cannot be set up a priori, and must be defined empirically.
In the following, we perform an heuristic search
for the optimal value of the regularization parameter to 
denoise a signal from the core collapse catalog of
\cite{Dimmelmeier:2008}. Since the procedure is the 
same regardless of the catalog, for the case of the BBH catalog
we only give at the end of this section 
the results of the corresponding search.
The goal is to find a small span of values of $\mu$ that provide 
a recovered (denoised) signal for all test signals
under different signal-to-noise ratio (SNR) conditions.
We shall apply the rROF algorithm for the time domain
and the SB method for the frequency domain.

\begin{figure}
  \centering
    \includegraphics[width=7.0cm]{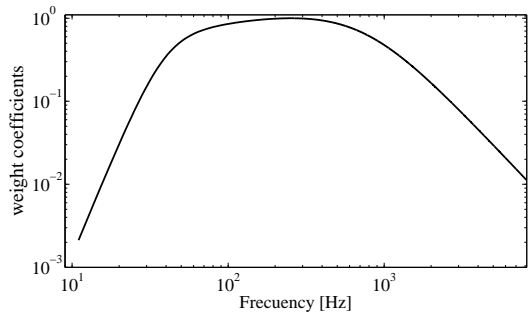}
  \caption{Frecuency distribution of noise weight coefficients obtained from the Advanced LIGO sensitivity curve.}
  \label{fig:weight_distribution}
\end{figure}

Standard algorithms assume stationary additive white Gaussian noise. 
However, as the noise of actual interferometric detectors is non-white, 
since the sensitivity is frequency dependent, 
we have to adapt the denoising algorithms to take this fact into account.
The weight distribution of the noise, ${w}$,      
according to the sensitivity curve of Advanced LIGO \cite{Schutz:2009}
is shown in Fig.~\ref{fig:weight_distribution}.

On the one hand, in the time domain we do not make any assumption 
about the noise, i.e., we use the rROF algorithm and we
filter out the obtained result below the lower cut-off frequency of the
sensitivity curve, according to the weight distribution.
On the other hand, in the frequency domain we proceed as follows: 
Given the observed signal $g$,
\begin{equation}
g=x+n\,,
\end{equation} 
where $x$ is the signal from the catalog and $n$ is the noise,
we compute the Fourier transform of the mirror extension of $g$, $gf$. Then, we 
solve the following TV-denoising model
\begin{equation}
\label{eq:rof_weigth}
v_{\rm opt}=\underset {{v}} {\text{argmin}}\int |\nabla{v}|+\frac{{\mu}}{2}||{v}-{gf}||_w^2~,
\end{equation}
where $||v ||_w^2 := \int w \cdot |v|^2$ is 
the weighted (by $w$) $L_2$-norm of $v$
in the frequency domain, by using the SB method for complex functions
of real variable. Finally we compute the inverse Fourier transform of $v_{\rm opt}$, and after 
restricting its values to the appropiate time domain, we obtain the denoised signal $u_{\rm opt}$.
We remark that due to its appropriate border treatment and computational efficiency,
we use the matrix formulation of the SB method developed
by Michelli et al (see  \cite{Micchelli:2011} for details) when we 
address intensive real-time calculation.

All signals of the core collapse catalog have been resampled
to the LIGO/Virgo sampling rate of $16,384$ Hz and
zero padded to be of equal length. 
For the SB algorithm in the frequency domain, 
signals have been Hanning windowed and mirror extended 
to avoid border effects in the Fourier Transform.
We add non-white Gaussian noise to the signals generated
as explained in Appendix \ref{noisegeneration}.
To ensure the best conditions for the convergence of the algorithms 
and to avoid round-off errors, we also scale the amplitude of the 
test signals $g$ of both catalogs to vary between -1 and 1.
The values of $\mu$ we discuss in this section are
hence determined by this normalization.
As we use the same noise frame in all the experiments
for comparison reasons, the amplitude of all signals
is scaled to the same values and differences are only
given by the SNR which for a wave
strain $h$ is defined as
\begin{equation}
\text{SNR}=\sqrt{4\Delta t^2 \Delta f \sum_{k=1}^{N_f}\frac{|\tilde{h}(f)|^2}{S(f_k)}}~,
\end{equation}
where $\tilde{h}$ indicates the Fourier transform of signal $h$ and $S(f_k)$ is the power spectral density.

\begin{table}
\caption{\label{tab:fidelity_tab} Values of the fidelity term and of the optimal value of $\mu$  for several time windows for the core collapse signals A, B, and C. The values are obtained after applying the SB algorithm in the frequency domain.}
\begin{ruledtabular}
\begin{tabular}{ccccc}
$\Delta_t ({\rm ms})$&$||g-x||^2_{L_2}$&
\multicolumn{3}{c}{$\mu_{\rm opt}$}\\
\cline{3-5}
& & A&B &C\\
\hline
1000&0.059 & 0.29&0.28&0.22\\
500&0.059& 0.45&0.49&0.36\\
250&0.059& 0.87&0.98&1.06\\
125&0.055& 1.11& 0.98&0.61\\
62.5&0.055& 1.45&1.66&2.80\\
31.25&0.055& 2.60&3.05&3.94\\

\end{tabular}
\end{ruledtabular}
\end{table}

The optimal value of the regularization parameter, 
$\mu_{\rm opt}$, is defined to be the one which 
gives the best results according to a suitable metric function
applied to the denoised signal and the original one, measuring the
quality of the recovered signal. 
In our case, we choose the peak signal-to-noise ratio, PSNR, 
based on the fidelity term, Eq.~(\ref{eq:fidelity_term}), 
\begin{eqnarray}
\label{eq:psnr}
{\rm PSNR(dB)}&=&10\log_{10}\left(\frac{N}{\text{MSE}}\right) \,,
\\
\text{MSE} &= &\frac{||x-u||_{L_2}^2}{N}~,
\end{eqnarray}
where  $x$ is the original signal from the catalog, $u$ is the
processed signal after applying the algorithms,
and $N$ is the number of samples. 

\begin{figure*}
        \centering
        \begin{subfigure}
        {\includegraphics[width=5.0cm]{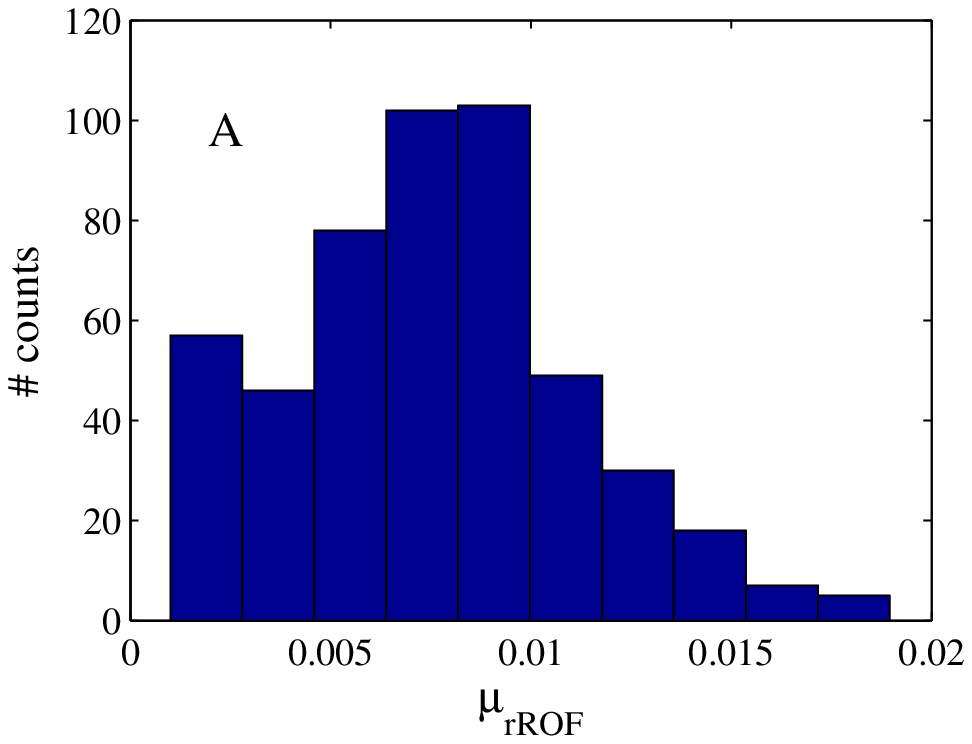}}
        \end{subfigure}
        \begin{subfigure}
    	{\includegraphics[width=5.0cm]{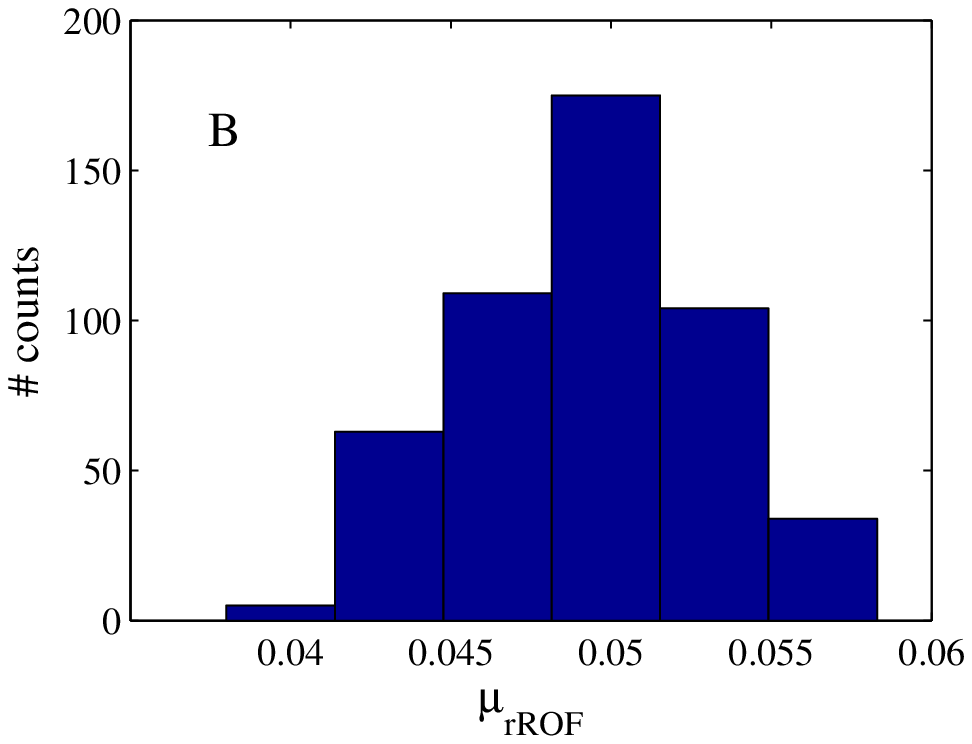}}
        \end{subfigure}
        \begin{subfigure}
    	{\includegraphics[width=5.0cm]{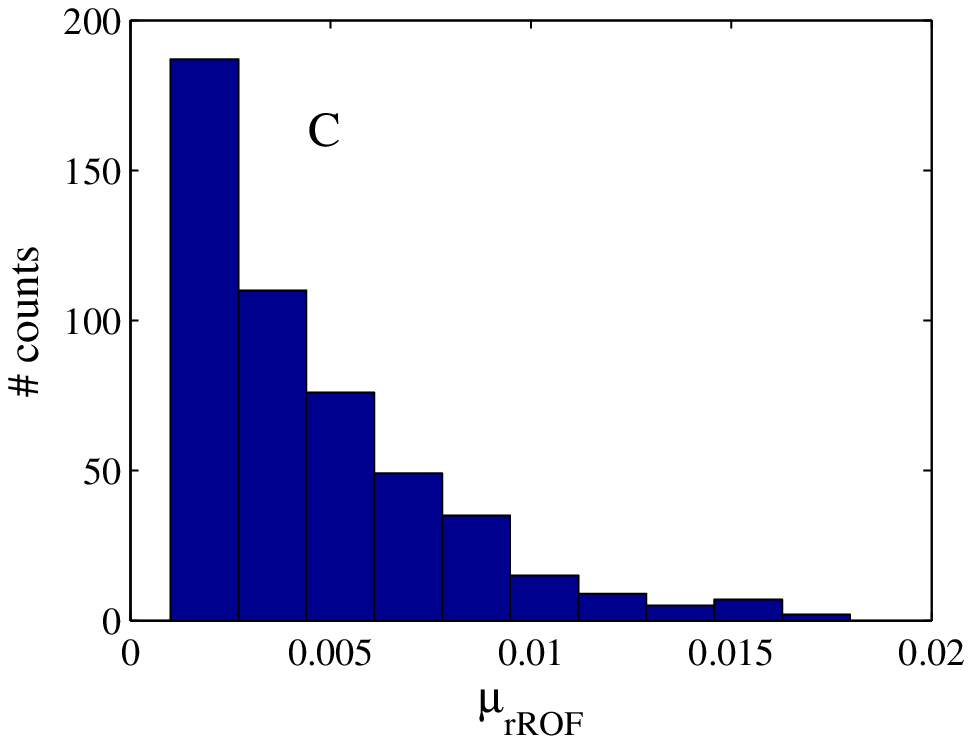}}\\
        \end{subfigure}
         \begin{subfigure}
    	{\includegraphics[width=5.0cm]{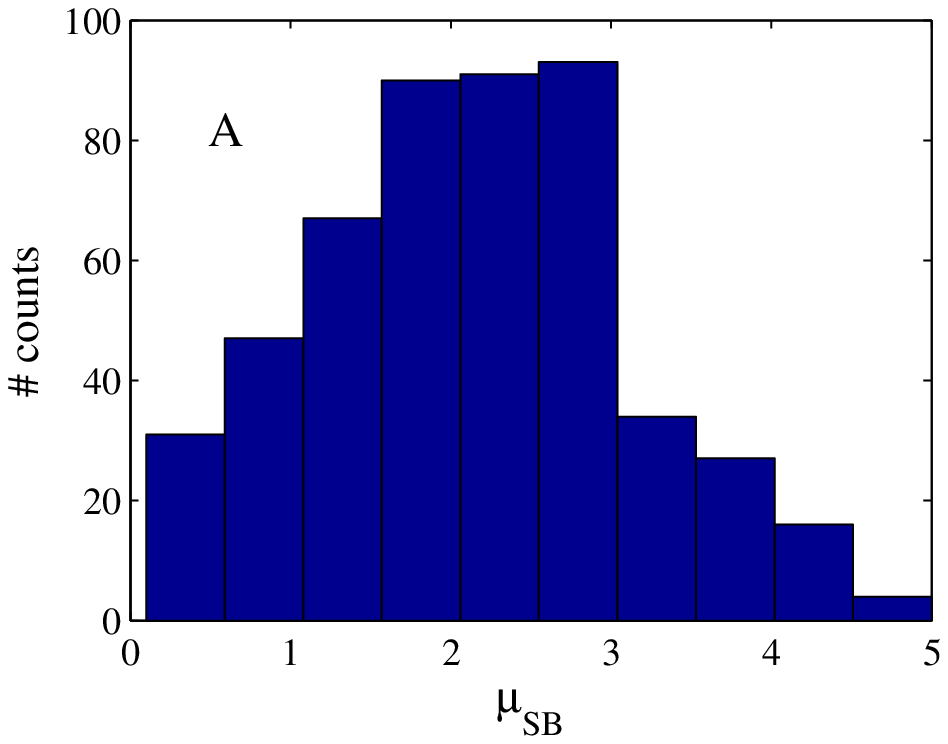}}
        \end{subfigure}
        \begin{subfigure}
    	{\includegraphics[width=5.0cm]{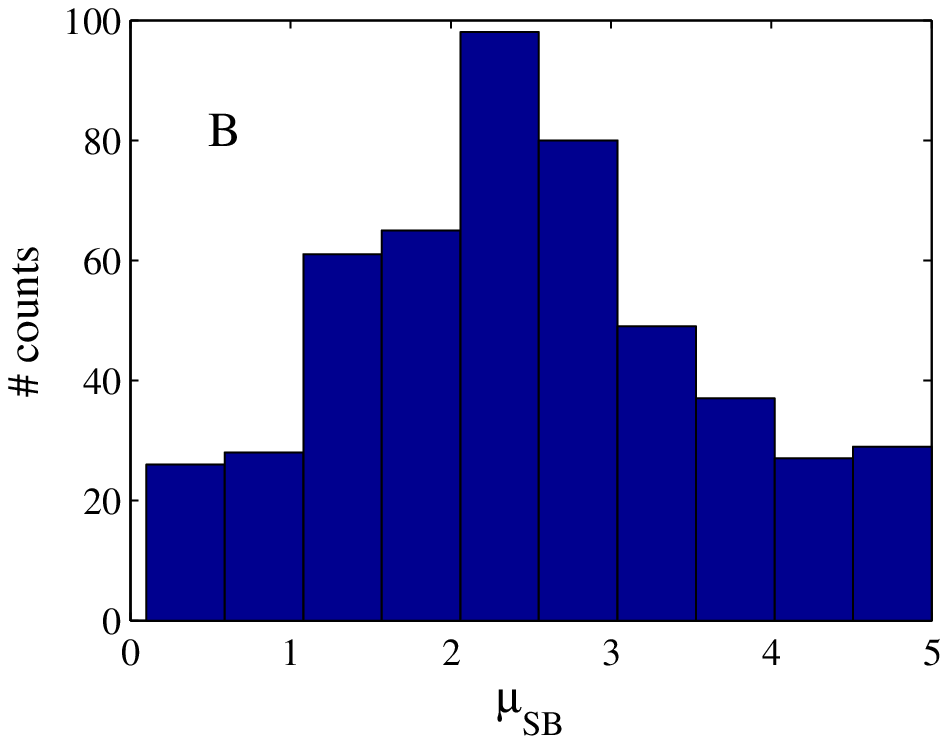}}
        \end{subfigure}
         \begin{subfigure}
    	{\includegraphics[width=5.0cm]{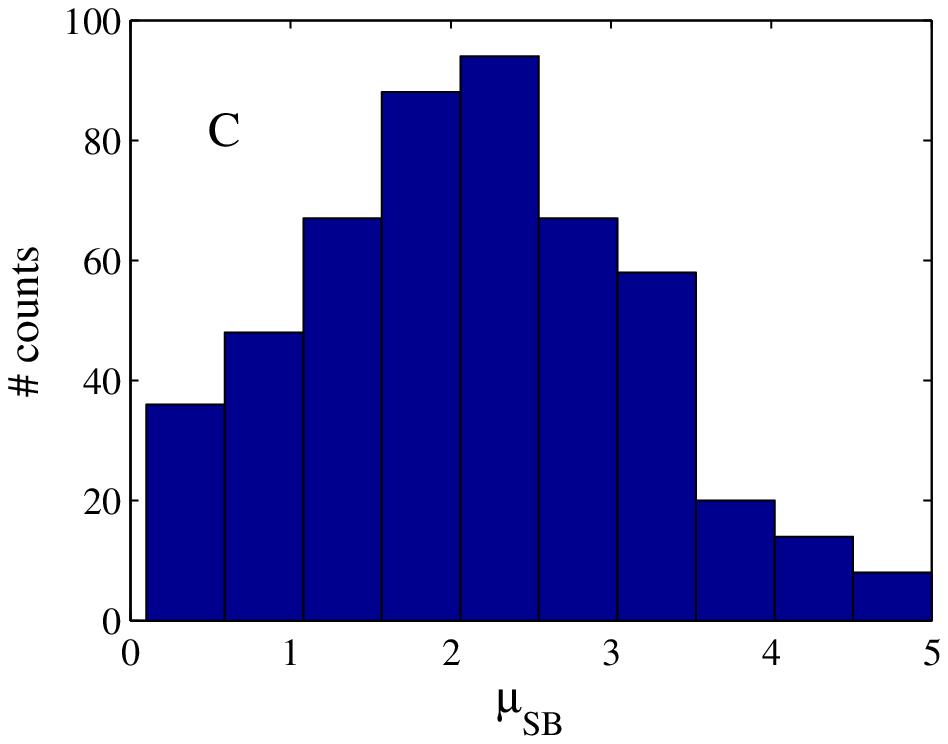}}
        \end{subfigure}
        \caption{Histograms of the values of $\mu_{\rm opt}$ for 500 noise generations for the three representative core collapse    
        signals. The upper panels show the values of $\mu_{\rm opt}$ for the rROF method in the time domain while the 
        corresponding results for the SB method applied in the frequency domain are shown in the bottom panel. A SNR=15 is 
        assumed.}\label{fig:histogram_psnr}
\end{figure*}

First of all, we have to find the appropriate time window to perform this comparison. 
Since bursts have short duration (a few ms), if the time window is 
long the signal to compare with is going to be composed of mainly zeros,
while, if it is short, some parts of the signal can be lost.
To study this dependence, we seek for the value of $\mu$ in
core collapse signals for which the fidelity term of the
denoised signal matches the fidelity term of the original signal,
that is, we seek for $\mu_{\rm opt}$ subject to 
$||g-x||_{L_2}^2\approx ||g-u||_{L_2}^2$.

\begin{figure*}
        \centering
         \begin{subfigure}
    	{\includegraphics[width=5.0cm]{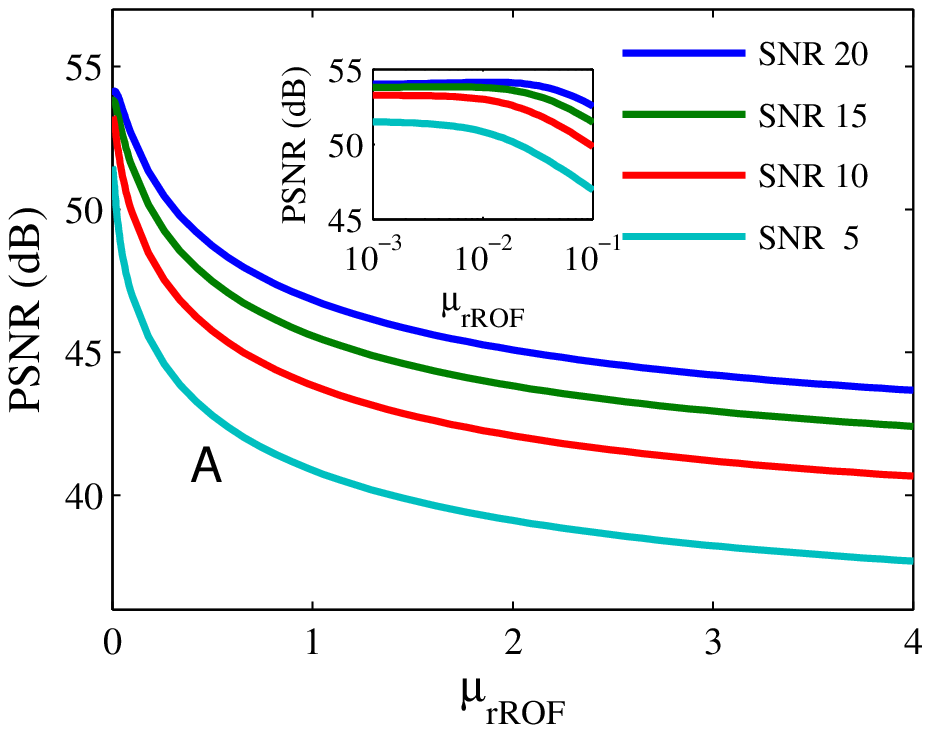}}
        \end{subfigure}
        \begin{subfigure}
    	{\includegraphics[width=5.0cm]{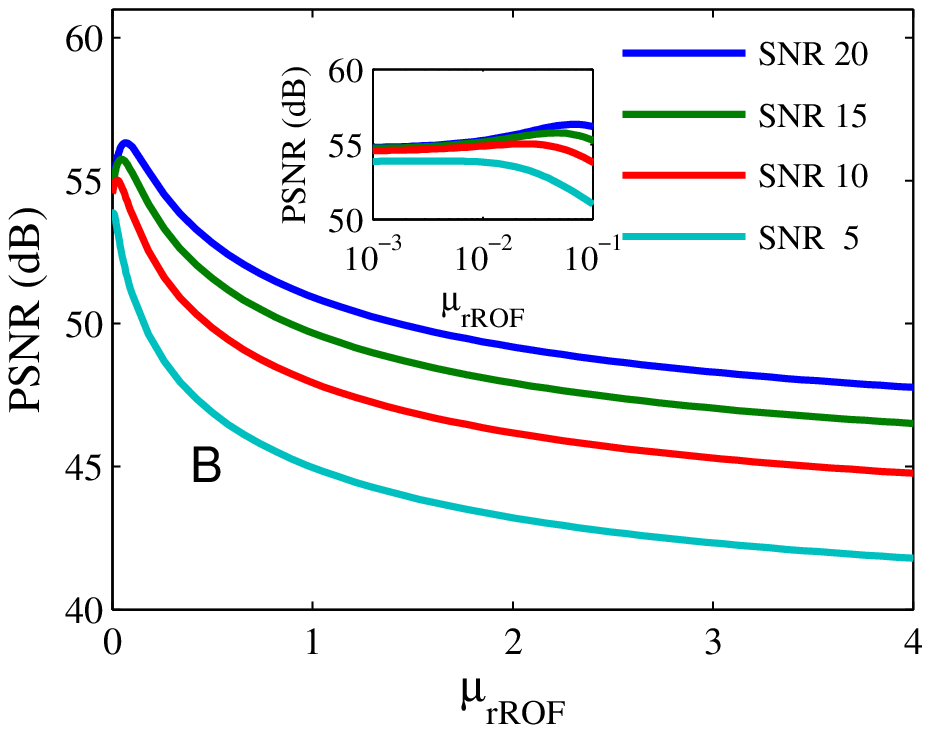}}
        \end{subfigure}
         \begin{subfigure}
    	{\includegraphics[width=5.0cm]{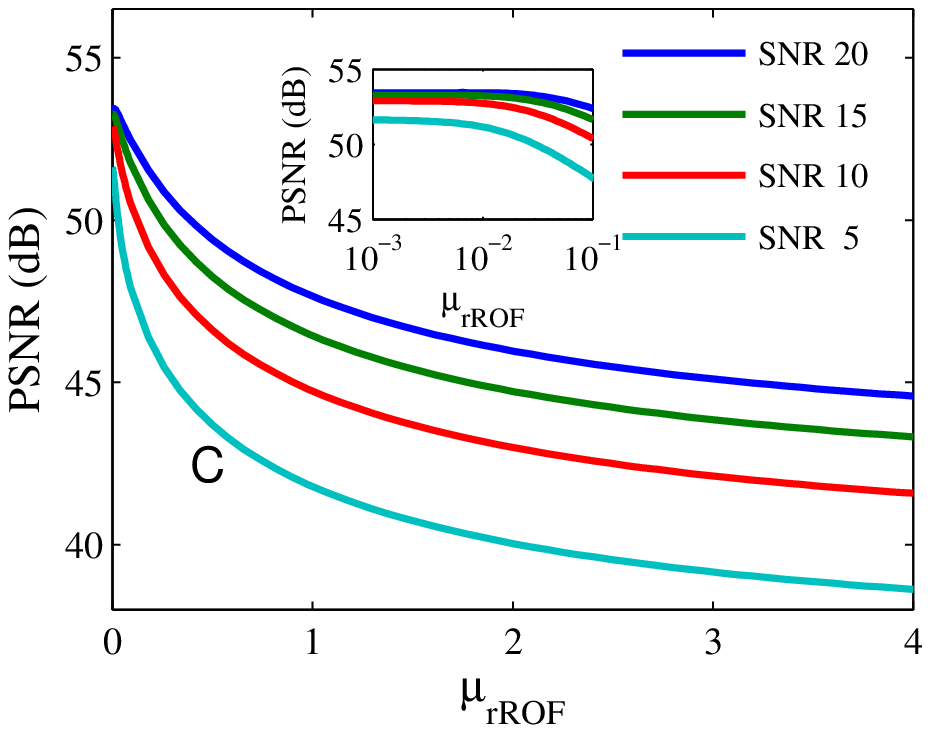}}\\
        \end{subfigure}
        \begin{subfigure}
        {\includegraphics[width=5.0cm]{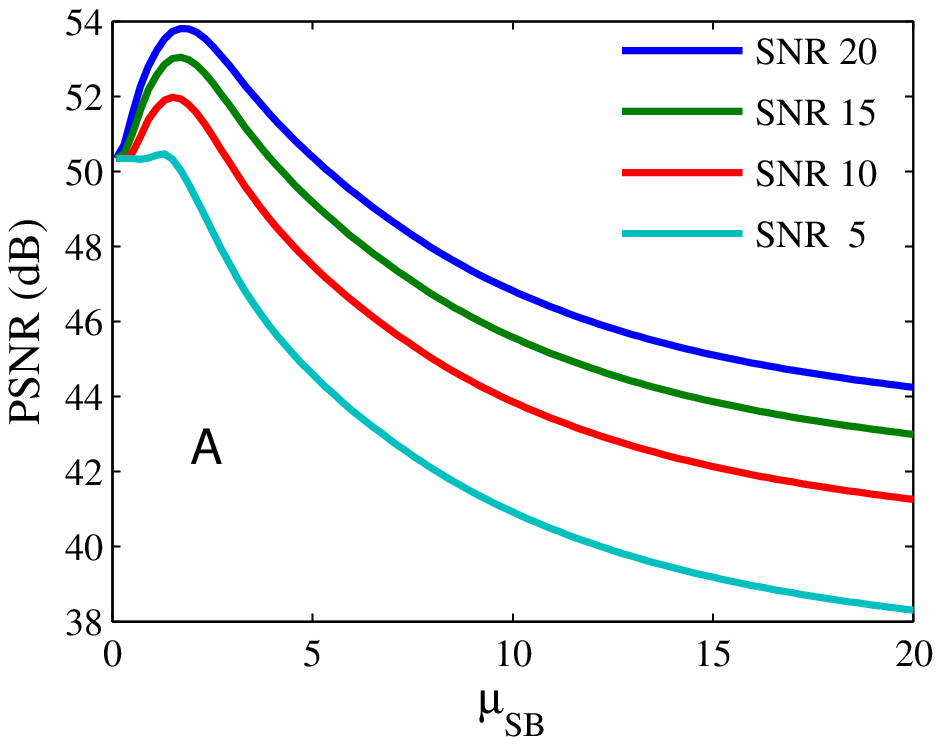}}
        \end{subfigure}
        \begin{subfigure}
    	{\includegraphics[width=5.0cm]{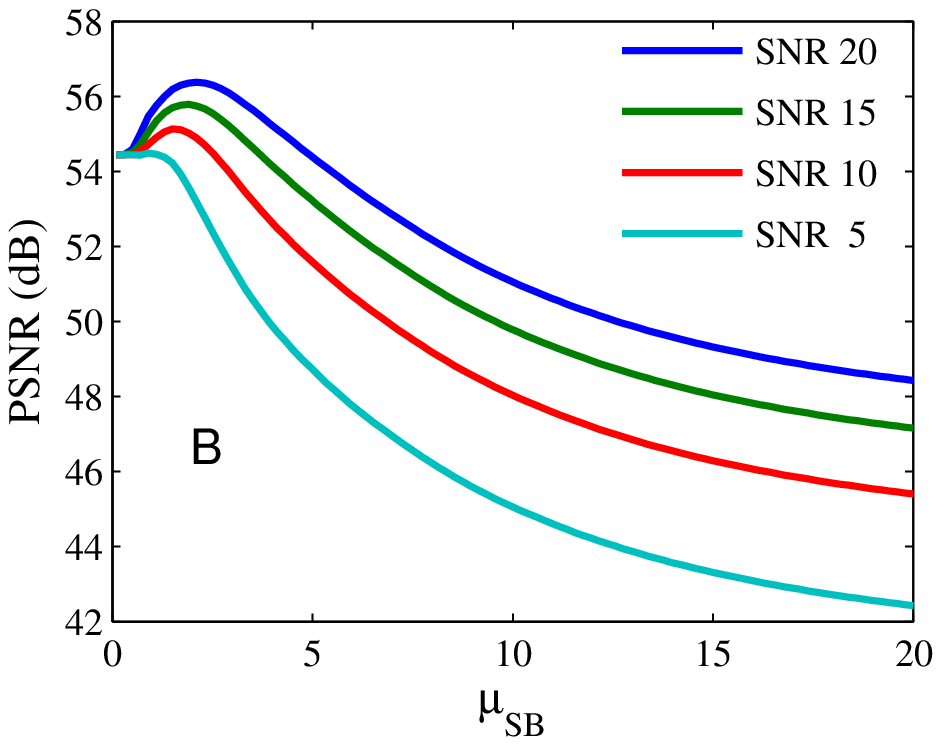}}
        \end{subfigure}
        \begin{subfigure}
    	{\includegraphics[width=5.0cm]{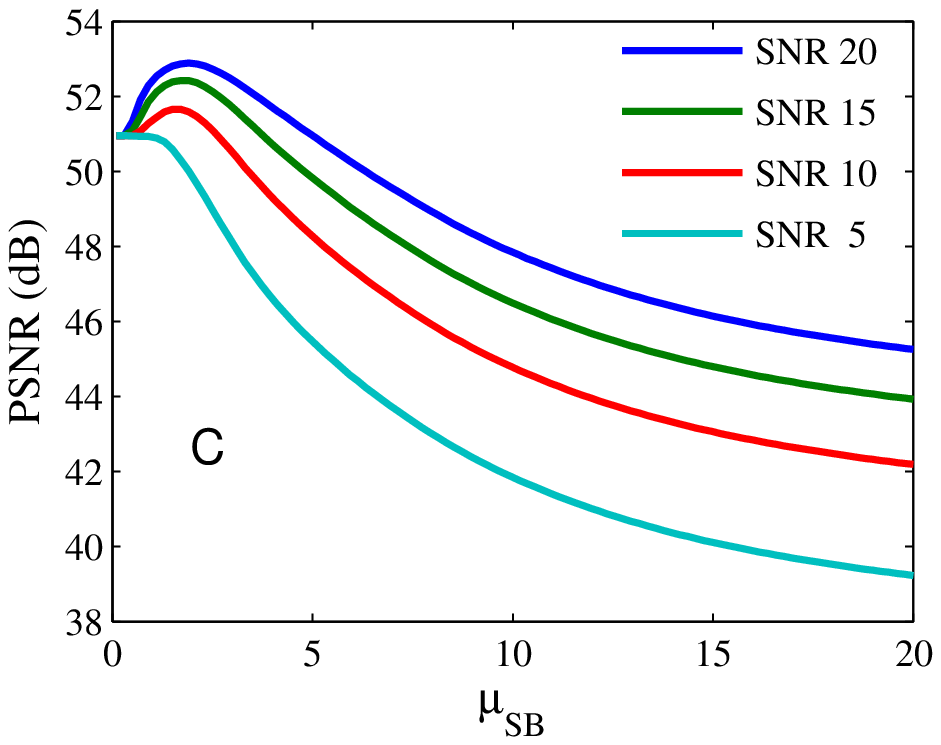}}
        \end{subfigure}
 	\caption{Dependence of the PSNR for different signal-to-noise ratios for the three representative core collapse signals.
	The upper panels show the values of PSNR for the rROF method in the time domain while the 
        corresponding results for the SB method applied in the frequency domain are shown in the bottom panel. The insets in the
        top panels magnify the areas where the variations in the curves are larger.}\label{fig:psnr_snr}
\end{figure*}

Results of this study for the three representative signals
from the core collapse catalog are shown in Table~\ref{tab:fidelity_tab}.
The SB method  has been explicitly developed for discrete signals and
it is well known that there exists a correlation between
the number of samples and the value of $\mu_{\rm opt}$.
Therefore, different (time or frequency) scales of the same signal
cannot be recovered using the same value of $\mu$.
Indeed, we find that the values of $\mu_{\rm opt}$ we obtain
show certain dependence on the number of samples, 
which is roughly equal to $\frac{1}{\sqrt{2}}$.
Both, differences in  $||g-x||^2_{L_2}$ and deviations
from the previous ratio are due to the weight that
the significant features of the GW signals have
relative to the number of zeros.
The rROF algorithm, on the other hand,
reduces the staircase effect associated to the shrinkage operator in SB.
The results of Table~\ref{tab:fidelity_tab} for the SB method
allow us to adjust the time window of the rROF method.
From this comparison we choose a time window of $62.5$ ms
for the core collapse signals as it yields a complete
representation of the waveforms without losing any significant feature.

Once we have selected the time window,
we must find the appropriate value of $\mu$ based on the PSNR value.
First we seek the optimal value of the regularization parameter
for several realizations of noise.
The corresponding histograms with the optimal value of $\mu$
for both algorithms, SB and rROF, are shown in
Fig.~\ref{fig:histogram_psnr} for the three representative burst signals,
assuming a SNR value of 15.
We note that both distributions have the expected Gaussian shape. 
The variance of each distribution gives us a window of variability
around the mean value of $\mu$ to estimate the optimal one.
The half-Gaussian in signal C for rROF is apparent and the whole Gaussian
can be seen by log-scaling the range of $\mu$.

Having found the mean value of $\mu$ for a given signal 
through noise variations, we extend the analysis to consider
different signal-to-noise ratios. 
We re-scale the amplitude of the three signals to fix the value of the SNR.
The results are displayed in Fig.~\ref{fig:psnr_snr} for both algorithms.
This figure shows that for all SNR values considered,
the PSNR values peak around the optimal value of the
regularization parameter.
The span of values of $\mu$ to ensure a proper denoising is 
$1.5-3$ for the SB method and $0.001-0.015$ for the rROF algorithm.
For very noisy signals (low SNR) the recovered ones 
are very oscillatory and cannot be distinguished from noise. 
In this case, it might be possible to apply 
other data analysis techniques
to improve the results. From our analysis we observe that in some cases
there is an interval of optimal values of $\mu$ giving the same
regularized result, probably due to the lack of resolution in the data.

\begin{figure}
 \centering
    \includegraphics[width=4.1cm]{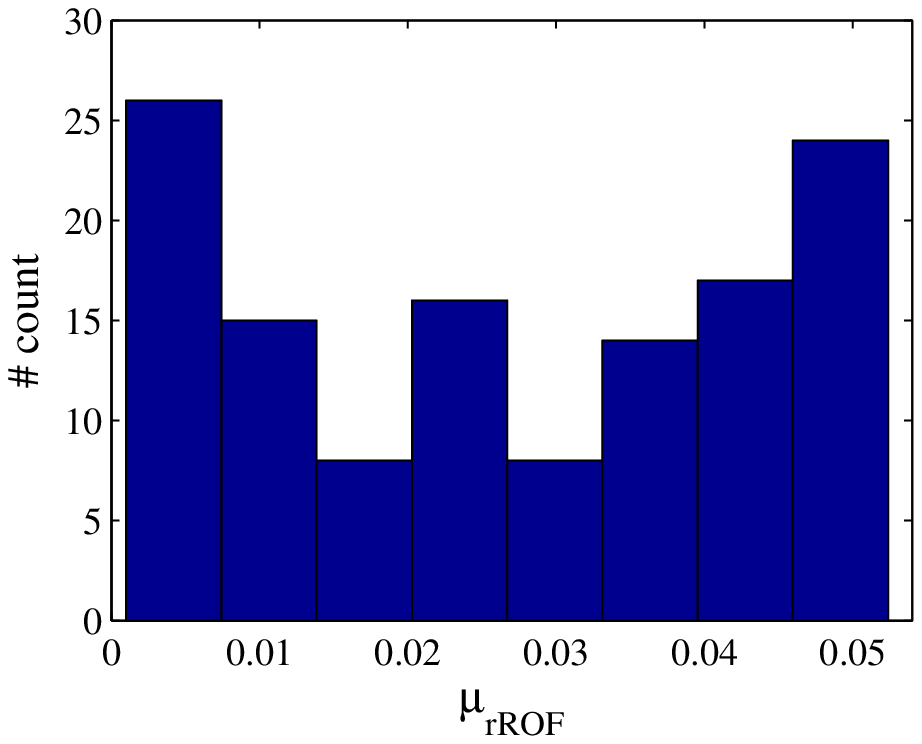}
     \includegraphics[width=4.2cm]{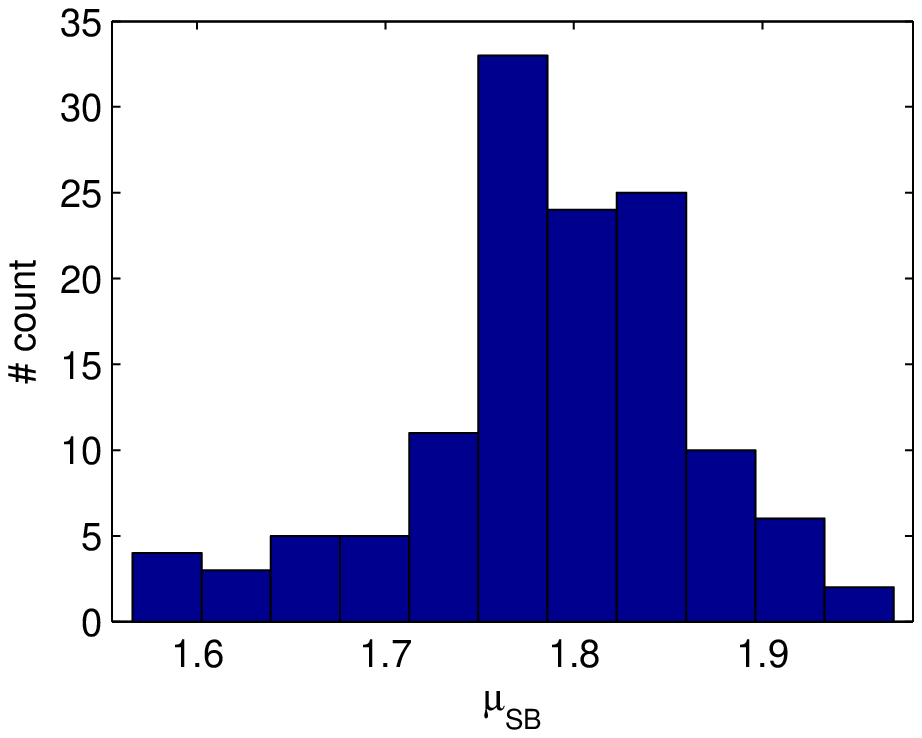} 
  \caption{Histograms of $\mu_{\rm opt}$ for all signals of the core collapse catalog 
  with $\text{SNR} = 15$. The left panel shows the span of values of $\mu_{\rm opt}$ for 
  the rROF method while the right panel displays the corresponding values for the SB method.}
   \label{fig:signal_dependence}  
\end{figure}

Finally, we analyze the optimal value for {\it all} 
signals of the core collapse catalog assuming SNR=15. 
As shown in Fig.~\ref{fig:signal_dependence}, 
the optimal value of $\mu$ is quite different 
across the entire catalog, particularly for the rROF algorithm (left panel). 
The values of $\mu_{\rm SB}$ span the interval $1.5-2$ 
and show a Gaussian profile, while for the case of the 
rROF algorithm they span the interval $0.001-0.05$ and 
do not show an obvious trend. 
We note that although the range of values of $\mu_{\rm opt}$ is large, 
it is still possible to perform the denoising procedure 
with acceptable results, by choosing the mean value 
for all the catalog signals and then tuning it up 
to find the value of $\mu$ that provides the best results. 

As mentioned at the beginning of this section
we have also performed the same analysis with
the entire catalog of BBH signals~\cite{Mroue:2013}.
In this case we choose a window that contains the 
last $12-14$ cycles before the merger ($\Delta t\sim 62.5$ ms) 
for illustrative reasons. 
The results obtained are similar to those reported for the 
core collapse catalog, the optimal interval for the rROF algorithm 
being $0.001-0.01$ and $1-3$ for the case of the SB algorithm.

 \begin{figure}
  \centering
    \includegraphics[width=7.0cm]{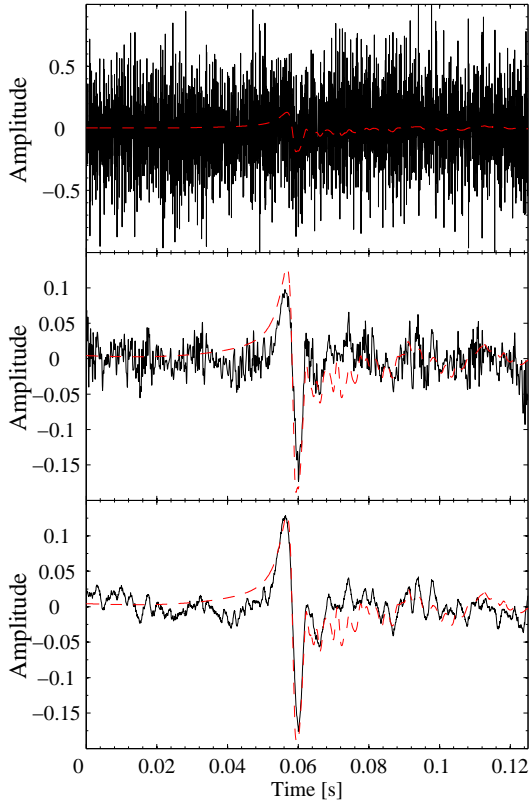}
  \caption{Denoising of the core collapse waveform signal C with $\text{SNR} = 20$. Top panel: original signal  (red dashed line) and non-white Gaussian noise (black solid line). Middle panel: Original and denoised (black solid line) signals for the SB method  in the frequency domain with $\mu=2.0$. Bottom panel: Original and denoised signals for the rROF method in the time domain with $\mu=0.09$.} 
    \label{fig:burst_denosing}
\end{figure}

We can conclude that the appropriate values of $\mu$ for both algorithms are restricted to a small 
enough interval which remains approximately constant for all signals of the catalogs and for different 
SNR. We stress that the concrete values of $\mu_{\rm opt}$ reported here are mainly 
used as a rough guide to apply the denoising procedures. If the properties of the signals change, 
such as the sampling frequency, the number of samples, or the noise distribution, 
the values of $\mu_{\rm opt}$ can also change, 
and it would become necessary to recompute them. 
Nevertheless, as we will show next, it is indeed 
possible to obtain acceptable results for all signals 
using a generic value of $\mu$ within the intervals 
discussed here which can then be fine-tuned 
to improve the final outcome.

\begin{figure}
        \centering
         \begin{subfigure}
        {\includegraphics[width=7.0cm]{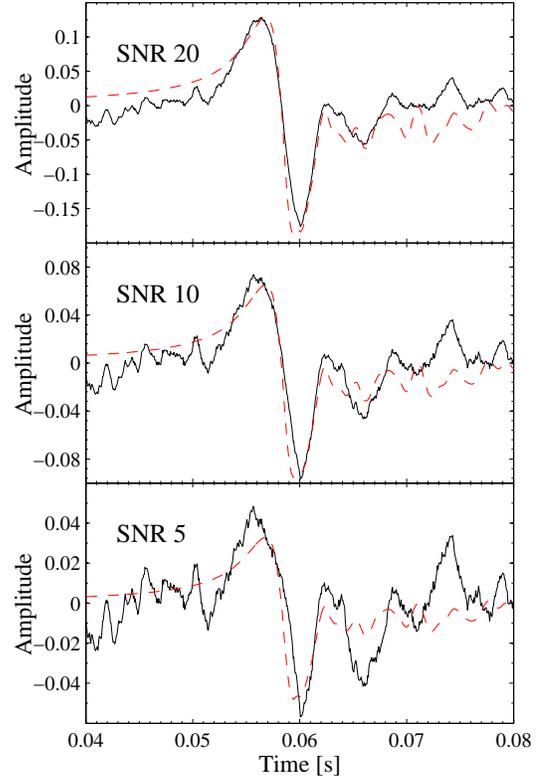}}
        \end{subfigure}
	\caption{Denoising of the core collapse waveform signal C for the rROF method with $\mu=0.09$ for three values of the SNR, 20 (top), 10 (middle), and 5 (bottom).}
	\label{fig:burst_denosing_SNR}
	\end{figure}

\section{Results}
\label{section:application}

\subsection{Signal Denoising}

We start applying both TV denoising methods to signals from both catalogs in a high SNR scenario, 
namely SNR=20. Our aim is to show how the two algorithms perform the denoising irrespective of the 
nature of the gravitational waveform considered. We assume that there is a signal in the dataset 
obtained from a list of candidate triggers, and that all glitches have been removed. This simple situation 
allows us to test our techniques as a denoising tool to extract the actual signal waveform from a noisy 
background. We apply the proposed methods, rROF in the time domain and SB in the frequency domain,
independently.

The results from applying our denoising procedure to a signal 
from the core collapse catalog is shown in the three panels of
Fig.~\ref{fig:burst_denosing}.
For the sake of illustration we focus on signal C, 
since the results are similar for the other two types of signals.
The most salient features of this signal are the two large positive 
and negative peaks around $t\sim 0.06$ s associated with 
the hydrodynamical bounce that follows the collapse of the iron core, 
and the subsequent series of small amplitude oscillations 
associated with the pulsations of the nascent proto-neutron star.
Note that contrary to Fig.~\ref{fig:signals} 
the time in this figure is not given with respect to the time of bounce. 
In the top panel we plot the original signal (red dashed line) 
embedded in additive non-white Gaussian noise (black solid line).
The middle panel shows the result of the denoising procedure 
after applying the SB method in the frequency domain with $\mu=2.0$,
and the bottom panel shows the corresponding result after applying 
the rROF algorithm in the time domain with $\mu=0.09$.
The two large peaks are properly captured and denoised,
most notably the main negative peak. 
This is expected due to the large amplitude of these two peaks,
as TV denoising methods work best for signals with a large gradient. 
In turn, those parts of the signal with small gradients 
cannot be recovered as nicely, as seen in the damped pulsations 
that follow the burst and which have amplitudes much smaller 
than the noise. We note that both algorithms attenuate positive and negatives 
peaks due to noise effects. If desired, it would be possible to  recover the actual 
amplitude of the main two peaks of the signal accurately by using a larger 
value of $\mu$. However, this would introduce a more oscillating 
signal in the part of the waveform with small gradients. 
Such oscillations are consistently more common for the SB method 
than for the rROF method, as can be seen from the middle and bottom 
panels of Fig.~\ref{fig:burst_denosing}.

The effect of varying the SNR on the denoising procedure is shown 
in Fig.~\ref{fig:burst_denosing_SNR} for the same core collapse waveform C. 
This figure magnifies the signal around the late collapse and early 
post-bounce phase, i.e. $0.04$s$\,<t<0.08$s. The three panels show, from 
top to bottom, the comparison between the denoised and the original signal
for SNR=20, 10, and 5, respectively. Only the results for the rROF algorithm 
with $\mu=0.09$ are shown, as the results and the trend found for the SB method 
are similar (only more oscillatory in the small gradient part of the signal 
for the latter). This figure shows that, as the SNR decreases,
the denoised signal recovers the original signal worse, as expected. 
While the amplitude of the oscillations of the denoised signal 
increases in the part with small gradients, it is nevertheless noticeable the 
correctness of the method to recover the amplitude of the largest negative 
peak of the signal even for SNR=5. We stress that all three signals have been 
denoised applying the same value of $\mu$ and recall that the study of the 
dependence of $\mu$ on the SNR (Section \ref{section:parameter_estimation}) 
predicts a lower value of $\mu_{\rm opt}$ as the SNR decreases to obtain the 
best results. Therefore, the effect of using a value of $\mu$ greater than the optimal 
one is also noticeable in the middle and lower panels of Fig.~\ref{fig:burst_denosing_SNR}.

\begin{figure}
  \centering
    \includegraphics[width=7.0cm]{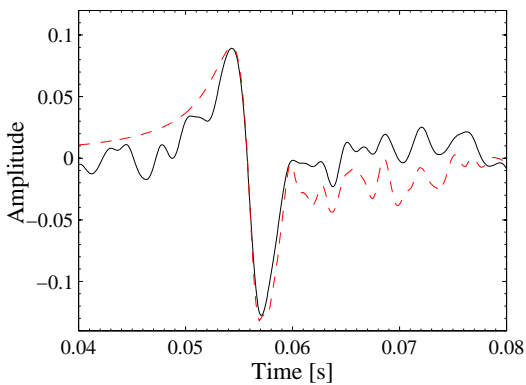}
  \caption{Denoising of the core collapse waveform signal C with $\text{SNR}=20$ after applying 
  both algorithms sequentially, rROF first with $\mu=0.05$ followed by SB with $\mu=8$.} 
  \label{fig:burst_denosing_tf}
\end{figure}

The amount and amplitude of the oscillations of the 
denoised signals in small gradient regions can be somehow made less
severe when both algorithms are applied sequentially. 
This is shown in Fig.~\ref{fig:burst_denosing_tf} 
for the GW burst signal C with SNR=20. 
The denoised signal in black in this figure is the result of 
applying an initial denoising with the rROF algorithm in a 1 s window 
followed  by a second step with the SB method in the frequency domain using
a 125 ms time window. The values of the regularization parameters
employed in this case are $\mu=0.05$ for the rROF algorithm and 
$\mu=8$ for the case of the SB method. 
We note that, in general, larger values of the 
regularization parameters are required when applying both methods 
sequentially because regularization is accumulative. 
Indeed, the output of the first step contains less noise and, therefore, the
second step needs larger values of $\mu$, i.e. less regularization.

 %
 \begin{figure}
  \centering
    \includegraphics[width=7.0cm]{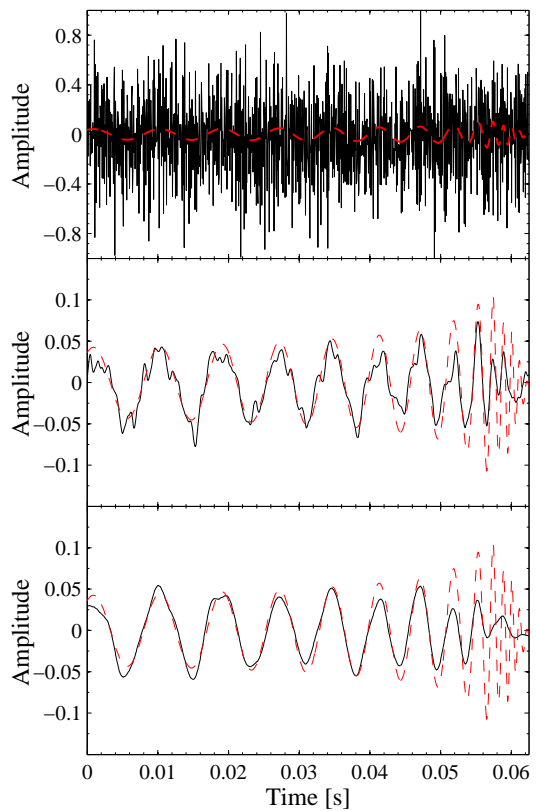}
\caption{Denoising of the BBH waveform signal ``0001" with $\text{SNR} = 20$. Top panel: original signal  (red dashed line) and non-white Gaussian noise (black solid line). Middle panel: Original and denoised (black solid line) signals for the SB method  in the frequency domain with $\mu=1.6$. Bottom panel: Original and denoised signals for the rROF method in the time domain with $\mu=0.0026$.}
    \label{fig:bbh_denosing}
\end{figure}

We turn next to apply the denoising procedure to the BBH catalog. 
For illustrative purposes we focus on a single signal of
such catalog, signal ``0001", as this suffices to reveal 
the general trends. The results are shown 
in Fig.~\ref{fig:bbh_denosing}. 
As before, the top panel shows the original signal (red dashed line) 
embedded in additive non-white Gaussian noise (black solid line). 
The middle panel shows the result of the denoising procedure after 
applying the SB method in the frequency domain with $\mu=1.6$, 
and the bottom panel shows the corresponding result after applying 
the rROF algorithm in the time domain with $\mu=0.0026$. 
A value of SNR=20 is assumed. 
To better visualize the results we display only the last few cycles 
up until the two black holes merge.

As BBH waveforms have longer durations than bursts and 
the characteristics of the signal change over time, 
scales between the beginning and the end of the signal 
are significantly different. 
This becomes clear in Fig.~\ref{fig:bbh_denosing} where
as a result of the scale variations in frequency and amplitude 
during the late inspiral and merge, the signal cannot be 
properly denoised throughout using the same value of $\mu$.
Typically we find that using a comparatively large value 
of $\mu$ helps to accurately recover larger amplitudes and 
high frequencies than lower frequencies and amplitudes, and vice versa. 
The reason for this is again due to the fact that the methods 
we employ are gradient dependent, preserving the large gradients 
and removing the small ones.  On the one hand, choosing a low value of 
$\mu$ to recover low frequency cycles makes the merger signal to be 
treated as high frequency noise. 
On the other hand, choosing a high value of $\mu$ to recover the 
merger part produces high oscillations in the rest of the signal. 
We have checked that it is nevertheless possible to obtain good 
results for the entire BBH waveform train using different values
of $\mu$ for different intervals of the waveform.

\subsection{Signal Detection}

From the previous analysis, it becomes manifest that in a low SNR situation our denoising algorithms alone cannot remove enough noise to produce detectable signals. In order to improve our results in such a situation, we can combine our techniques with the use of spectrograms. Such an approach is usually employed in GW data analysis to seek for transient power peaks in the data that could correspond to actual  gravitational wave signals, assuming that all known transients have been removed from the data  \cite{Parameswaran:2014,credico:2005, anderson:1999}.

\begin{figure}
 \centering
    \includegraphics[width=7.5cm]{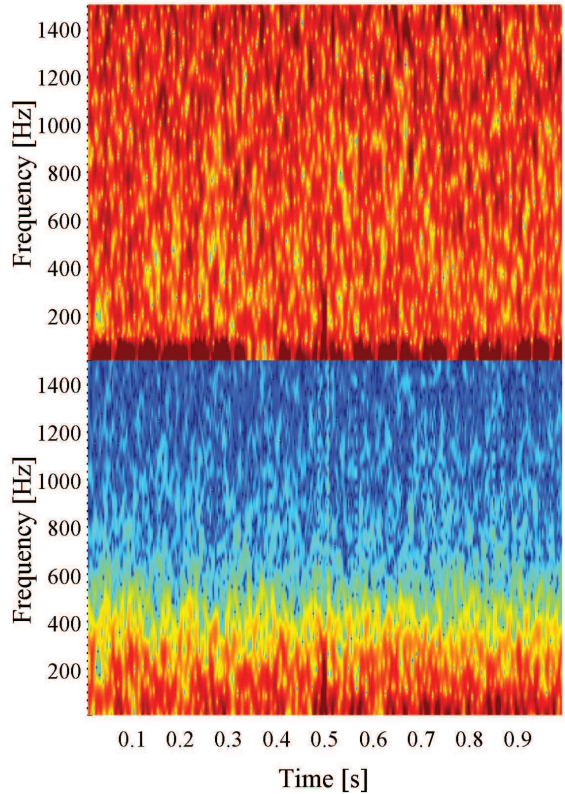}
 \caption{Spectrograms of the original noisy signal (top panel) and denoised signal (bottom panel) for the core collapse signal C and SNR =10. The higher values of the spectral power density are shown in red, while the lower power is represented in blue.}
   \label{fig:spectrogram_comparison}  
\end{figure}

First, we have to check if the information we can obtain from the spectrogram would be modified by the application of our techniques. To do this we compare the spectrogram from an original noisy signal with SNR=10 with the spectrogram of the corresponding denoised signal. The results are shown in Fig. \ref{fig:spectrogram_comparison} for the core collapse signal C. The spectrogram of the original signal is shown in the top panel and the denoised spectrogram is shown in the bottom panel. Red color represents high spectral power while lower power is represented in blue.  In order to produce Fig.~\ref{fig:spectrogram_comparison} we first apply the rROF algorithm in the time domain followed by the SB method in the frequency domain, both for a 1s time window, and then we calculate the spectrogram. The power peak around 0.5s corresponds to the GW signal which is clearly distinguishable in both spectrograms, and its structure remains similar after the denoising procedure. 

\begin{figure}
  \centering
    \includegraphics[width=7.5cm]{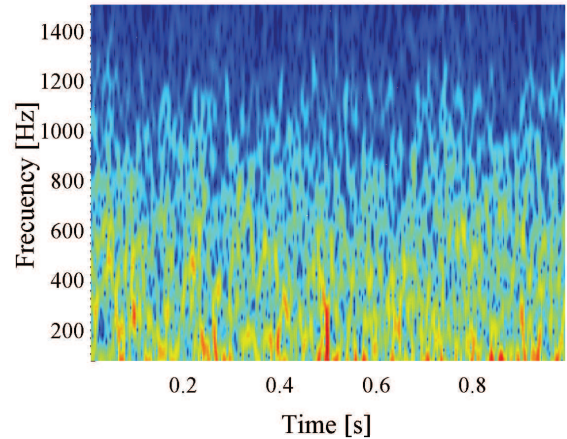}
  \caption{Spectrogram of the core collapse signal C for SNR=5. The excess power around $0.5$ s is supposed to be produced by the gravitational wave signal.}
   \label{fig:spectrogram}
\end{figure}

An example of the application of the spectrogram together with our algorithms in a low SNR situation where the spectrogram alone would not reveal any high power peak, is shown in Fig.~\ref{fig:spectrogram}. This figure displays the denoised spectrogram of the same core collapse signal C originally embedded in non-white Gaussian noise but now with SNR=5. In this case we have a dataset which contains 1s of data from the detector. The exact  arrival time of the signal is unknown and is what we want to determine by computing the spectrogram. In order to find the time of arrival of the signal we integrate the power of the first 2000 Hz for each temporal channel and look for the channel that contains the maximum power. After selecting this channel we perform the denoising procedure only in this channel, using first the SB method as a filter and then applying the rROF algorithm in order to obtain the signal waveform.  Fig.~\ref{fig:spectrogram} shows that the power peak of the signal (red color) is clearly distinguishable from the noisy background (green-blue colors). This findings give us confidence to use our algorithms as a denoising tool jointly with other data analysis techniques.

\section{Summary and outlook}
\label{section:summary}

In this paper we have presented new methods and algorithms for denoising gravitational wave signals.
The methods we use are based on $L_1$ norm minimization  and have been originally developed and fully tested in
the context of image processing where they have been shown to be the best approach to solve the so-called
Rudin-Osher-Fatemi denoising model. We have applied these algorithms to two different types of numerically-simulated gravitational wave signals, namely bursts produced from the core collapse of rotating stars and chirp waveforms from
binary black hole mergers. The algorithms have been applied in both the time and the frequency domain. Both of
our methods, SB and rROF, reduce the variation of the signal, assuming that due to its randomness the larger
variations are due to the noise. We have performed an heuristic search to find the set of values best suited for
denoising gravitational wave signals and have applied the methods to detect signals in a low signal-to-noise ratio
scenario without any a priori information on the waveform. In particular, the rROF algorithm in the time
domain has led to satisfactory results without  any assumption about the noise distribution. On the other hand,
in order to apply the SB method in the frequency domain, we have selected a particular weight distribution.
This distribution can be chosen freely so as to adjust it to the specific spectral characteristics of the noise or of the
detector sensitivity curve. Overall, we conclude that the techniques we have presented in this paper may be used
along with other common techniques in gravitational wave data analysis (e.g.~spectrograms) to help increase the
chances of detection. Likewise, these methods should also be useful to improve the results of other
data analysis approaches such as Bayesian inference or matched filtering when used as a noise removal initial
step that might induce more accurate results for the aforementioned traditional methods.

In future work we plan to further test these TV-based methods in a more realistic setup by using
real noise from the detectors instead of the somewhat simplistic non-white Gaussian noise employed in this paper.
The presence of transient glitches may be a crucial factor spoiling the results as our
methods cannot discriminate if the denoised signals correspond to a real signal or a glitch or outlier.
There is room to improve the capabilities of the methods by incorporating
information about the sources into the algorithms through the use of
signal dictionaries from numerical relativity waveform catalogs and known glitches,
and through the implementation of a machine learning
algorithm to both, improve the pure denoising results and
remove spurious information.

\emph{Acknowledgments.} It is a pleasure to thank I.S.~Heng and A.~Sintes for helpful discussions. This work was supported
by the Spanish MICINN (AYA 2010-21097-C03-01), MINECO (MTM2011-28043) and by the Generalitat Valenciana (PROMETEO-2009-103).	

\appendix
\section{Noise Generation}
\label{noisegeneration}

We generate non-white Gaussian noise whose shape corresponds to Advanced LIGO in the proposed broadband configuration. For this purpose, we employ the algorithm libraries (LAL)  \cite{url:LAL} provided by the
LIGO Scientific Collaboration (LSC) that include one-sided detector noise power spectral density $\sqrt{S(f)}$.
If necessary, random noise series can be obtained weighting samples from Normal distribution by noise power spectral density,
\begin{eqnarray}
\mathbb{R}\text{e}(\tilde{n}(f))&=&\frac{\sqrt{S(f)}}{2} \text{N}(\mu,\sigma^2_f)~,
\\
\mathbb{I}\text{m}(\tilde{n}(f))&=&\frac{\sqrt{S(f}}{2} \text{N}(\mu,\sigma^2_f)~,
\end{eqnarray}
\begin{eqnarray}
\{\tilde{n}(f)\in \mathbb{C}~:~f=0, \Delta_f,2 \Delta_f,...,(N-1)\Delta_f \} ,
\end{eqnarray}
where $\Delta_f=\frac{F_s}{N}$ being $F_s$ the sampling rate and $N$ the number of samples. Mean $(\mu)$ and variance $(\sigma_f^2)$ are set to 0 and 1 respectively as corresponds to the standard normal distribution.

As noise time-series are real-valued, in the frequency domain noise must satisfy
\begin{equation}
\tilde{n}(-f)=\tilde{n}^*(f)\,,
\end{equation}
and
\begin{equation}
\mathbb{I}\text{m}\{\tilde{n}(0)\}=\mathbb{I}\text{m}\{\tilde{n}(f_{Ny})\}=\mathbb{I}\text{m}\{\tilde{n}(-f_{Ny})\}=0~,
\end{equation}
where $f_{Ny}=\frac{F_s}{2}$ corresponds to the Nyquist frequency.

Properties of the Fourier transform assure that the 
Gaussian character of the noise signal 
is preserved when changing domain.
Therefore, we obtain noise time-series 
applying the discrete Fourier Transform (DFT)
\begin{eqnarray}
\tilde{n}(f_k)&=&\sum_{j=0}^{N-1}n(t_j) e^{\frac{-2\pi i}{N}kj}~,
\\
n(t_j)&=&\frac{1}{N}\sum_{k=0}^{N-1}N(f_k) e^{\frac{2\pi i}{N}kj}~.
\end{eqnarray}



\end{document}